\documentclass[preprint2]{aastex}

\slugcomment{submitted to Astronomical Journal}

\shorttitle{The Star Formation in Normal Galaxies}
\shortauthors{Boselli et al.}

\begin{document}


\title{1.65~$\rm \bf \mu$m (H-band) surface photometry of galaxies. 
VI: The history of star formation in normal late-type galaxies}


\author{A. Boselli\altaffilmark{}}
\affil{Laboratoire d'Astrophysique de Marseille, Traverse du Siphon, F-13376 Marseille
Cedex 12, France}
\email{Alessandro.Boselli@astrsp-mrs.fr}

\author{G. Gavazzi\altaffilmark{}}
\affil{Universit\`a degli Studi di Milano - Bicocca, P.zza dell'Ateneo Nuovo 1, 20126 Milano, Italy}
\email{Giuseppe.Gavazzi@uni.mi.astro.it}

\author{J. Donas\altaffilmark{}}
\affil{Laboratoire d'Astrophysique de Marseille, Traverse du Siphon, F-13376 Marseille
Cedex 12, France}
\email{Jose.Donas@astrsp-mrs.fr}

\and

\author{M. Scodeggio\altaffilmark{}}
\affil{IFCTR-CNR, via Bassini 15, Milano, Italy}
\email{marcos@ifctr.mi.cnr.it}




\begin{abstract}
We have collected a large body of NIR ($H$ band), 
UV (2000 \AA) ~and H$\alpha$ measurements 
of late-type galaxies. These are used,
jointly with spectral evolutionary synthesis models, to study the initial mass 
function (IMF) in the mass range $m$ $>$ 2 ${\rm M\odot}$. For spirals (Sa-Sd),
Magellanic irregulars (Im) and blue compact dwarfs (BCD), 
our determination is consistent with
a Salpeter IMF with an upper mass cutoff $M_{up}$ $\sim$ 80 ${\rm M\odot}$.
\noindent
The history of star formation and the amount of total gas (per unit mass) 
of galaxies are found to depend primarily on their total masses 
(as traced by the 
$H$ band luminosities) and only secondarily on morphological type.
The present star formation activity of massive spirals
is up to 100 times smaller than that average over their lifetime, 
while in low mass galaxies 
it is comparable to or higher than that at earlier epochs.
Dwarf galaxies have presently larger gas reservoirs per unit mass
than massive spirals. 
\noindent
The efficiency in transforming gas into stars and the time scale for gas 
depletion ($\sim$ 10 Gyrs) are independent of the 
luminosity and/or of the morphological type. 
\noindent
These evidences are consistent with the idea that galaxies are coeval systems, 
that they evolved as closed-boxes 
forming stars following a simple, universal star
formation law whose characteristic time scale is small ($\tau$ $\sim$ 1 Gyr)
in massive spirals and large ($\tau$ $>$ 10 Gyr) in low mass galaxies.
A  similar conclusion was drawn by Gavazzi \& Scodeggio (1996) to explain
the colour-magnitude relation of late-type galaxies.
\noindent
The consequences of this interpretation on the evolution of 
the star formation rate and of the gas density 
per comoving volume of the Universe with look-back time are discussed.
\end{abstract}


\keywords{Galaxies: general -- Galaxies: evolution -- Galaxies: spiral -- Galaxies: ISM 
-- Stars: formation --Ultraviolet: galaxies}


\section{Introduction}
 
The hierarchical (Cole et al. 1994) and the monolithic collapse (Sandage 1986) 
scenarios of galaxy formation make different 
predictions on the evolution of the primeval density perturbations 
which gave birth to galaxies. While in hierarchical models small objects
form first and large objects are formed via successive merging of 
smaller structures, in the monolithic scenario galaxies are formed via 
a unique collapse of the original density perturbation without any
interaction with the environment (closed box). 
A wide variety of physical and morphological properties observed 
in nearby galaxies are expected from both models: while in a monolithic 
scenario ellipticals should have formed most of their stars in a time 
short compared to the collapse time in order to avoid the formation of a disc
by gas-gas dissipation, in spirals the slower initial star formation activity
allowed a significant fraction of gas to form a rotating disc (Sandage 1986).
In hierarchical models the different properties of galaxies
are the result of the initial conditions of the original merger.    

The predicted
luminosity functions and density distributions of galaxies in the Universe
should in principle allow to disentangle observationally between these two 
models
(Kauffmann et al. 1999). 
Furthermore these models make predictions
on the time evolution of the physical properties of galaxies, such as 
their structural parameters (e.g. bulge-to-disk ratio), their dynamics,
their star formation history (thus their stellar populations), 
their gaseous content and metal enrichement
(Kauffmann \& Charlot 1998 and references therein). 
An accurate determination of the dependence of these properties on $z$
(e.g. Cowie et al. 1996) will allow us to discriminate 
between the galaxy formation scenarios.\\
A detailed knowledge of the phenomenology of galaxies at $z$=0, 
which represents the present stage of galaxy 
evolution, is of primary importance for constraining evolutionary models.\\
Ironically the "zero point" of galaxy evolution is still not sufficiently 
well determined, due to the lack of extensive observations of
local galaxies. For example the systematics of their current
star formation activity, gaseous content and metallicity 
are still poorly known.
In particular the UV and H$\alpha$ observations which are representative of
the current star formation activity of galaxies (Donas et al. 1987, 
Kennicutt 1998), the near-IR data which trace the old stellar population,
and spectro-photometric measurements (Zaritsky et al. 1994) which are 
necessary to derive the
properties of the underlying stellar population and the physical 
conditions of the interstellar medium, are still rare.\\
With the ambitious aim of constructing a representative description
of the physical properties of nearby galaxies, we
undertook a multi-frequency survey of $\sim$ 3500 optically selected galaxies
spanning the broadest possible range in morphological type (Ellipticals, 
Spirals, dE, 
Im and BCD), luminosity (-22 $\leq$ $M_B$ $\leq$ -13), and environmental 
conditions (clusters - isolated).
Our own H$\alpha$,  
near-IR, millimetric and centimetric observations, together with data
from the literature, were collected in a multifrequency database
(see Gavazzi, Boselli \& Pierini, 1996).\\
This database has been used to interpret the colour-magnitude relation 
(Gavazzi \& Scodeggio 1996) and  the star formation activity of late-type 
galaxies 
(Gavazzi et al. 1998). 
The extension of our near-IR survey to faint spirals, Im and and blue compact
dwarfs (BCD), 
which is the subject of the present series of papers, combined with
recent H$\alpha$ observations of a set of late-type galaxies,
and with the large body of available HI and CO radio data
makes it possible to study, in the present work, the phenomenology of the 
young stellar population and of 
the gas content of a large sample of late type galaxies from Sa to Im-BCD.\\

\noindent
Previous works devoted at analysing the phenomenology of nearby galaxies
have shown that the fraction of available gas and the activity of star 
formation (normalized to the visible stellar luminosity) increases 
along the Hubble sequence (Roberts \& Haynes 1994; 
Kennicutt 1998). However none of these works was focused on the role
of the total mass in governing the evolution of galaxies, an issue
that we address in the present work. We do so using the observation 
that the near-IR $H$ luminosity, which is dominated by the emission of the 
old main sequence and red giant 
stars (Bruzual \& Charlot 1993), is a direct tracer of the dynamical mass
of late-type systems (within the optical radius; Gavazzi et al. 1996a).
Since the near-IR mass to light ratio was found not to vary with mass for galaxies
spanning the whole range in morphological type from Sa to Im and BCD,
contrary to what was found in the optical, 
by normalizing other observed quantities (e.g. the gas content etc.) to
the near-IR luminosity we better remove than using optical luminosities
the well known observed luminosity-luminosity or luminosity-mass scaling 
relations. Furthermore the decomposition of the near-IR light profiles
are used to discriminate between disc dominated and bulge dominated galaxies.
This new approach is made possible by the availability of a large
body of near-IR observations of nearby galaxies described
in papers I, II, III and IV of this series. Furthermore the available
HI and CO data allow a precise determination of the 
total gas content for a sample of galaxies with a broad coverage
in morphological type and luminosity.
 
The sample used and the data coverage is described in Section 2.
Early-type (type $\le$ S0a) galaxies are excluded from the present analysis
since their UV emission, being dominated 
by the extreme horizontal branch stars and 
their progenitors (Dorman et al. 1995), does not give information on the young
stellar population. Moreover
we restrict the present analysis to galaxies which do not show
signs of interaction with their surroundings. 
For this we use the HI-deficiency parameter
to discriminate between "normal" galaxies and galaxies suffering for
gas depletion due to ram pressure and we include in the present study
only isolated objects or cluster galaxies with a HI-deficiency 
(defined as the ratio of the HI mass to the 
average HI mass of isolated objects
of similar morphological type and linear size (Haynes \& Giovanelli 1984))
$\leq 0.3$.\\
The current star formation rate (SFR) and the initial mass function (IMF)
of late-type galaxies are studied in Section 3. 
The dependence of the total (atomic plus molecular) gas content as a 
function of morphological type and luminosity is analysed in Section 4. 
A model for the star formation history of late-type galaxies is
presented in Section 5.
The dependence of the SFR and of the gas density of the Universe on 
look-back time is briefly discussed in Section 6.
An appendix describes how the H$\alpha$ and UV (2000 \AA) data are 
corrected for [NII] contamination and internal extinction. 

\section {The sample}

Galaxies analysed in this work are 
taken from the Zwicky catalogue 
(CGCG, Zwicky et al. 1961-1968)($m_{pg}$ $\leq$ 
15.7). They are either late-type (type$>$S0a) 
members of 3 nearby ($vel$ $\le$ 8000 km s$^{-1}$) clusters 
(Cancer, A1367, Coma), or located in the relatively low-density regions of 
the Coma-A1367 supercluster 
(11$^h$30$^m$ $<$RA$<$ 13$^h$30$^m$; 
18$^o$$<$dec$<$32$^o$) as defined in Gavazzi et al. (1999a).
To extend the present study to lower luminosities,
we include in the sample the late-type Virgo cluster galaxies brighter 
than $m_{pg}$ $\leq$ 14.0 listed in the 
Virgo Cluster Catalogue as cluster members (VCC, Binggeli et al. 1985).
Furthermore VCC galaxies with
14.0 $\leq$ $m_{pg}$ $\leq$ 16.0 included in the "ISO" subsample 
described in Boselli et al. (1997a) and CGCG galaxies 
in the region 12$^h$ $<$RA$<$ 13$^h$; 
0$^o$$<$dec$<$18$^o$ but outside the VCC, are considered.
To avoid systematic environmental effects
we consider the subsample of late-type galaxies whose HI-deficiency
is $\leq$0.3, typic of unperturbed, isolated galaxies.
The final combined sample comprises 233, mainly ``normal''
galaxies (a few starburst or active galaxies might however be included).
\\
The accuracy of the morphological classification is excellent for the 
Virgo galaxies (Binggeli et al. 1985; 1993).
Because of the higher distance, the morphology of galaxies belonging 
to the other surveyed regions suffers from an uncertainty 
of about 1.5 Hubble type bins.\\
We assume a distance of 17 Mpc for the members (and possible members) 
of Virgo cluster A, 22 Mpc for 
Virgo cluster B, 32 Mpc for
objects in the M and W clouds (see Gavazzi et al. 1999b).
Members of the Cancer, Coma and A1367 clusters are assumed at 
distances of 62.6, 86.6 and 92 Mpc respectively.  
Isolated galaxies in the Coma supercluster are assumed 
at their redshift distance adopting $H_o$ = 75 km s$^{-1}$ Mpc$^{-1}$.\\
For the 233 optically selected galaxies,
complementary data are available in other bands as follows:
100\% have HI (1420 MHz), 99\% $H$ band (1.65 $\mu$m) data available. 
A much sparser coverage 
exists in the UV (2000 \AA)(29 \%), CO (115 GHz)(38\%) and H$\alpha$ (6563 \AA)
(65\%), as shown in Table 1 and 2. 

The morphological distribution of galaxies is given in Table 3.

\subsection {Data analysis}

H$\alpha$+[NII] fluxes obtained from imaging, 
aperture photometry or integrated spectra 
are taken from Kennicutt \& Kent (1983), Kennicutt et al. (1984), 
Gavazzi et al. (1991) Gavazzi et al. (1998), Young et al. (1996)
Almoznino \& Brosch (1998), Moss et al. (1998), Heller et al. (1999) 
and references therein. H$\alpha$ fluxes from Kennicutt \& Kent (1983), 
Kennicutt et al. (1984) and Gavazzi et al. (1991) have been multiplied by 
1.16, as suggested by Kennicutt et al. (1994), in order to account
for the continuum flux overestimate due to inclusion of the telluric 
absorption band near 6900 \AA ~in the comparison filter.
Additional observations of 66 galaxies have been recently obtained by us 
during several runs at the Observatoire de Haute Provence (France), at San Pedro
Martir (Mexico) and at Calar Alto (Boselli et al. in preparation; Gavazzi et 
al. in preparation). The estimated error on the H$\alpha$+[NII] flux is 
$\sim$ 15\%.

The UV data (at 2000 \AA) are taken from the FOCA 
(Milliard et al. 1991) 
and FAUST (Lampton et al. 1990) experiments; UV
magnitudes are from Deharveng et al. (1994) and Donas et al. (1987;
1990; 1995; in preparation). The estimated error on the UV 
magnitude is 
0.3 mag in general, but it ranges from 0.2 mag for bright galaxies to 0.5 
mag for weak sources observed in frames with larger than average calibration 
uncertainties.

HI data are taken from Scodeggio \& Gavazzi (1993) and  
Hoffman et al. (1996, and references therein).
HI fluxes are transformed into neutral hydrogen masses
with an uncertainty of $\sim$ 10\%.  

CO data, used to estimate the molecular hydrogen 
content, are from Boselli et al. (1997b), Boselli et al. (1995)
and references therein. The average error on CO fluxes is 
$\sim$ 20\%; the error on the H$_2$ content, however, 
is significantly larger (and difficult to quantify) due
to the poorly known CO to H$_2$ conversion factor (see Boselli et al. 1997b).

NIR data, mostly from Nicmos3 observations, are taken from this series 
of papers: Gavazzi et al. (1996b,c: Paper I and II), Boselli et al. (1997a), 
Gavazzi et al. (2000a, Paper III), Boselli et al. (2000, Paper IV).
From these data we derive total (extrapolated to infinity)
magnitudes $H_T$ determined as described in Gavazzi et al. (2000b: Paper V)
with typical uncertainties of $\sim$ 10 \%. These are 
converted into total luminosities 
using: $log L_H = 11.36 - 0.4H_T +2logD$ 
(in solar units), 
where $D$ is the distance to the source (in Mpc). For a few objects we
derive the $H$ luminosity
from $K$ band measurements assuming an average $H-K$ colour of
0.25 mag (independent of type; see Paper III).

A minority of the objects in our sample have an $H$ band magnitude
obtained from aperture photometry, thus with no asymptotic extrapolation.
For these we use the magnitude $H_{25}$ determined as in Gavazzi \& Boselli 
(1996) at the optical radius
(the radius at which the $B$ surface brightness is 25 mag arcsec$^{-2}$)
which is on average 0.1 magnitudes fainter than $H_T$ (Gavazzi et al. 2000a,b).

The total $H$ magnitudes are corrected 
for internal extinction according to Gavazzi \& Boselli (1996).
No correction has been applied to galaxies of type $>$ Scd.
The model independent near-IR concentration index parameter $C_{31}$,
defined as the ratio of the radii containing 75 \% to 25 \% of the total 
light of a galaxy, will be used throughout the paper to discriminate 
between disc dominated and bulge dominated galaxies. As shown in paper V, 
pure exponential discs are characterized by $C_{31}$ $<$ 3,
while $C_{31}$ is $>$ 3 in galaxies with prominent bulges.

\section {The SFR in galaxies}

Various indicators of the star formation activity of late-type galaxies have 
been proposed in the literature (see for a review Kennicutt 1998).
In the present analysis we use the H$\alpha$ and UV (2000 \AA) luminosities, 
which are commonly accepted as the most direct indicators of the star 
formation rate (SFR).

\subsection {SFR from H$\alpha$ and UV (2000 \AA) luminosities}

The H$\alpha$ luminosity gives a direct measure of the global 
photoionization rate of the 
interstellar medium due to high mass ($m$ $>$ 10 ${\rm M\odot}$), young 
($\le$ 10$^7$ years) O-B stars (Kennicutt 1983; 1990; 
Kennicutt et al. 1994).
The total SFR can be determined by extrapolating the high-mass SFR to
lower mass stars using an assumed initial mass function (IMF) $\psi(m)$:

\begin{equation}
{\psi (m)=\int^{M_{up}}_{M_{low}} k~m^{-\alpha}dm }
\end {equation}

\noindent
where $M_{up}$ ($M_{low}$) is the upper (lower) mass cutoff and 
$\alpha$ the slope of the IMF.
In the assumption that the SFR is constant on a time scale of some 
10$^7$ years, 
the SFR (in ${\rm M\odot yr^{-1}}$) is given by the relation:

\begin{equation}
{SFR_{H\alpha} ({\rm M\odot yr^{-1}}) = K_{H\alpha}(\alpha,M_{up},M_{low}) 
L_{H\alpha} 
({\rm erg~s^{-1}})}
\end {equation}

\noindent
where $K_{H\alpha}(\alpha,M_{up},M_{low})$ is the 
proportionality constant between the 
H$\alpha$ luminosity $L_{H\alpha}$ (in erg s$^{-1}$) and the 
$SFR_{H\alpha}$ 
(in ${\rm M\odot yr^{-1}}$). The value of 
$K_{H\alpha}(\alpha,M_{up},M_{low})$,
which depends on the slope $\alpha$ and on the upper and lower mass cutoffs 
$M_{up}$ and $M_{low}$ of the IMF,
can be determined from models of stellar population synthesis.
Different values of $K_{H\alpha}(\alpha,M_{up})$ from the stellar 
population
synthesis models of Charlot \& Fall (1993) are given in Table 4,
all for a lower mass cutoff $M_{low}$ = 0.1 ${\rm M\odot}$ and a solar metallicity. 

The UV emission of a galaxy at 2000 \AA~ is dominated by the emission of
less recent ($\sim$ 10$^8$ years) and massive (2$<$ $m$ $<$ 5 ${\rm M\odot}$)
A stars (Lequeux 1988). The UV emission becomes stationary if the SFR 
is constant over an
interval of time as long as the life time of the emitting stars
on the main sequence, i.e. $\geq$ 3 10$^8$ years. Thus, assuming a
star formation rate constant on time scales of some 10$^8$ years, 
the rate of star formation
$SFR_{UV}$ from UV luminosities at 2000 \AA~ ($L_{UV}$) can be determined 
from the relation:

\begin{equation}
{SFR_{UV} ({\rm M\odot yr^{-1}}) = K_{UV}(\alpha,M_{up},M_{low}) L_{UV} ({\rm erg~s^{-1}A^{-1}})}
\end {equation}

\noindent
where $K_{UV}(\alpha,M_{up},M_{low})$ is the proportionality constant 
between the 
UV luminosity $L_{UV}$ (in erg s$^{-1}$\AA$^{-1}$) and the $SFR_{UV}$ 
(in ${\rm M\odot yr^{-1}}$). Values of $K_{UV}(\alpha,M_{up})$ 
at 2000 \AA ~ for $M_{low}$ = 0.1 ${\rm M\odot}$ and a solar metallicity kindly made
available by S. Charlot are listed in Table 4.

H$\alpha$ and UV fluxes can thus be used independently to estimate
the rate of star formation in galaxies once they are corrected for internal 
extinction and for the 
contribution of the [NII] emission line ($\lambda$ 6548 \AA~ and 6584 \AA) 
which contaminates the narrow band H$\alpha$ photometry. 
The corrections applied to the present data are described in the Appendix.
As discussed in the Appendix, the lack of far-IR data and of integrated 
spectra for all the sample galaxies prevents us to make accurate 
[NII] contamination and extinction correction for each single galaxy, forcing
us to use ``statistical'' corrections. Given the broad distribution in 
the UV and H$\alpha$ extinctions measured by Buat \& Xu (1996)
and given in Fig. 11 for each morphological class, we estimate an uncertainty
in the extinction correction of a factor of $\sim$ three for each single
galaxy, but probably lower for an entire class of objects. 
This uncertainty, which
might be responsible for part of the observed scatter in several
star formation indicators, such as the birthrate parameter, should not be
critical in the forecoming analysis since it is 
significantely smaller than the whole range in star formation observed 
in the sample galaxies, which spans more than 3 orders of magnitude.\\
From now on we estimate $SFRs$ (in ${\rm M\odot yr^{-1}}$) 
from UV and H$\alpha$
luminosities using $K_{UV}(\alpha,M_{up},M_{low})$ and
$K_{H\alpha}(\alpha,M_{up},M_{low})$ from Table 4 for
$\alpha$=2.5, $M_{up}$=80 ${\rm M\odot}$ and $M_{low}$=0.1 ${\rm M\odot}$; 
for galaxies with both UV and H$\alpha$ measurements 
(52 objects), the adopted 
$SFR$ is an average of the results of the two methods.

\subsection {The IMF}

Since the H$\alpha$ emission is due to massive ($m$ $\geq$ 10 ${\rm M\odot}$) 
O-B stars, and the UV emission to moderate mass 
(2 $\leq$ $m$ $\leq$ 5 ${\rm M\odot}$)
A stars, the two SFR determinations provide us with an indirect method for 
studying the IMF of spiral galaxies in the mass
range $m$ $\geq$ 2 ${\rm M\odot}$ (Buat et al. 1987; Bell \& Kennicutt 2000). 
This method can be applied
only if the SFR has been constant over the last
3 10$^8$ years, corresponding to the life time of stars dominating
the UV emission on the main sequence. It is well known that this assumption
applies for "normal" unperturbed objects, such as the ones  
selected here, while it does not hold for interacting systems 
(Kennicutt et al. 1987) and starburst galaxies.

For all galaxies in our sample with available H$\alpha$ and UV data 
(52 objects),
we compare in Fig. 1 the ratio of the H$\alpha$ to UV fluxes,
corrected for extinction and [NII] contamination as described in 
the appendix, with the values determined from
the stellar population synthesis models given in Table 4. 
Different symbols are use for disc dominated (filled circles)
and bulge dominated (empty circles) galaxies.
The logarithm of the H$\alpha$ to UV flux ratio is plotted
versus the morphological type and the $H$ luminosity.

The average value 
$log [flux H\alpha/flux UV(2000\AA)]$=1.43 $\pm$0.25 
is consistent with 
a IMF with slope $\alpha$ = 2.5 and $M_{up}$ = 80 ${\rm M\odot}$, 
and differs significantly from the values obtained for an IMF 
with slopes
1.5 or 3.5 (see Table 4). Figure 1 a and b show that 
$log [flux H\alpha/flux UV(2000\AA)]$
is independent of the morphological type, mass and of the presence of a bulge.
The value $\alpha$ = 2.5 is consistent with
the Salpeter IMF ($\alpha$ = 2.35; Salpeter 1955). 
As shown in Fig. 1, the H$\alpha$ to UV flux ratio is
more sensitive to the slope of the IMF than to $M_{up}$.

As discussed by Calzetti (1999), the H$\alpha$ to UV flux ratio should
increase with extinction in any fixed geometry. In the case of
a slab model (absorbing dust and emitting stars well mixed in a disc), the 
observed $H\alpha/UV(2800 \AA)$ flux ratio overestimates the
real value by a factor of 2
for typical H$\alpha$ extinctions of $\sim$ 1.1 mag (Calzetti 1999).
This value has however to be taken as un upper limit since it has
been observed that, despite a factor of $\sim$ 4 in the galactic
extinction law between 6563 \AA ~and 2000 \AA, UV and H$\alpha$
extinctions are comparable because of a less efficient 
absorption of the 2000 \AA ~ photons dictated by the less concentrated
distribution of the UV emitting stars relative to the ionizing stars 
inside the dusty HII regions (see Appendix).  
We expect that variations in the extinction are responsible for 
an important fraction of the scatter in the H$\alpha$ to UV flux ratio observed
in Fig. 1 without introducing any important second order systematic effect
(see Appendix).

An $\alpha$=2.5 is in agreement with the value obtained by Buat et al. 
(1987), who analysed
31 galaxies with H$\alpha$ and UV data, adopting a method similar 
to the one used in this work, coupled with an older version of the 
evolutionary synthesis 
models. We do not confirm the trend they find with the morphological type,
which might be contributed to by HI-deficient cluster galaxies
which we deliberately excluded from our analysis.

\section {The star formation history of galaxies}

\subsection{The birthrate parameter b}

The birthrate parameter $b$ is
defined by Kennicutt et al. (1994) as the ratio of the current SFR to the 
average past SFR. This distance independent parameter $b$ is given by:

\begin{equation}
{b=\frac{SFR}{<SFR>_{past}}=\frac{SFR ~t_o ~(1-R)}{M_{star}}}
\end{equation}

\noindent
where $t_o$ and $M_{star}$ are the age and the total stellar mass
of the disc and $R$ is the fraction of gas that stars re-injected 
through stellar winds into the interstellar medium during their lifetime.
While the SFR comes directly from the H$\alpha$ and the UV luminosities,
the remaining parameters must be estimated indirectly.  
Since stars eject different fractions of gas into the ISM at different
epochs of their life, the return parameter $R$ is a function of time 
and depends on the assumed IMF and birthrate history. 
Kennicutt et al. (1994) have shown that
90\% of the returned gas is released in the first Gyr (more than half
in the first 200 Myr) of a stellar generation for any assumed IMF. 
For this reason we can safely 
assume an instant recycling gas approximation in the determination of 
the birthrate parameter $b$, with a constant value $R$=0.3, as determined
by Kennicutt et al. (1994) for a Salpeter IMF.

The total stellar mass $M_{star}$ can be determined using the same 
procedure described in Kennicutt et al. (1994), but using $H$ band 
luminosities instead of optical ones.
Near-IR luminosities are directly proportional to
the total dynamical masses ($M_{tot}/L_H$=4.6) at the $B$ band
25 mag arcsec$^{-2}$ isophotal radius, 
independent of morphological type (Gavazzi et al. 1996a). 
The same $M_{tot}$ vs. $L_H$ relationship is
followed by BCDs, where however the near-IR flux might be contaminated by
the emission of red supergiants; thus even in these objects 
the $H$ luminosity can be properly used as an estimator of the total mass.

Assuming $t_o$ $\sim$12 Gyrs, the birthrate parameter $b$
comes directly from:

\begin{equation}
{b=\frac{SFR ~t_o(1-R)}{L_H(M_{tot}/L_H)DM_{cont}}}
\end{equation}

\noindent
where $DM_{cont}$ is the dark matter contribution to the $M_{tot}/L_H$
ratio at the optical radius, that we assume to be
$DM_{cont}$=0.5, as in Kennicutt et al. (1994). As discussed by Rubin (1987),
this value is independent on type and luminosity and thus it should not
introduce any systematic effect in the determination of $b$.
Sandage (1986) and Kennicutt et al. (1994) have shown that the birthrate
parameter, when determined using $B$ luminosities, increases along the Hubble
sequence, but it is still not known how $b$ scales with mass.

The birthrate parameter (in logarithmic units) is plotted in Fig. 2 a and b 
versus the morphological type and the $H$ luminosity.
Fig. 2 shows a strong relationship between the birthrate parameter and these
structural parameters, with late-type and dwarf galaxies (low
mass objects) all characterized by a similar present and past star 
formation rate
($b$ $\sim$ 1). There are a few galaxies with $b$ $>$ 1, i.e. with 
present star 
formation rate higher than averaged over the past. These are low-mass, 
dwarf galaxies which might undergo episodes of star bursts.
Massive spirals, on the contrary, have present SFRs significantly
lower than in the past ($b$ $\sim$ 0.1-0.01).
The relationship between $b$ and the $H$ luminosity is however considerably
clearer than with the Hubble type, whilst only a systematic difference
is observed between types $\leq$ Sbc ($b$ $\leq$ 0.25) and types
$\geq$ Scd ($b$ $\geq$ 0.25), with Sc spanning the whole range in $b$. 

If we limit our analysis to Virgo cluster galaxies, for which the 
morphological classification is more accurate, we observe a less 
dispersed relation between $b$ and the morphological type. A relationship
between $b$ and the $H$ luminosity is however still observed 
inside any given morphological class, implying that part of the
observed scatter in the $b$ vs. type relationship is due to
a stronger dependence of $b$ on the $H$ luminosity.

On the other hand part of the scatter in the $b$ vs. $log(L_H$) relationship
is due to the presence of a bulge; at any luminosity,
galaxies with strong bulges ($C_{31}$$>$3; open circles) have lower 
birthrate parameters than pure exponential discs ($C_{31}$$<$3; filled 
circles). 

\subsection {The gas content of galaxies}

It is well known that the fraction of atomic gas increases 
along the Hubble sequence (Roberts \& Haynes 1994, and references therein), but
it is still unknown how the total gas mass (atomic plus molecular) scales
with the total mass of galaxies.
The line emission of HI at 21 cm and of $^{12}$CO(1-0) at 2.6 mm
can be used to estimate the total gas content (HI + H$_2$) of the 
target galaxies.

The molecular hydrogen content can be estimated assuming a constant ratio 
between the
$^{12}$CO(1-0) line emission and the H$_2$ surface density. 
In this work we follow Boselli et al. (1997b), 
but adopting the CO to H$_2$ conversion factor 
$X$=1.0 10$^{20}$ mol cm$^{-2}$ 
(K km s$^{-1}$)$^{-1}$ of Digel et al. (1996). For galaxies with no 
CO measurement, we assume 
that the molecular hydrogen content is 10\% of the HI, as estimated 
from isolated
spiral galaxies by Boselli et al. (1997b). The total gas mass is
increased by a 30\% to take into account the helium contribution.
The total gas content (normalized to the mass) of a galaxy depends strongly 
on the morphological type and on the $H$ luminosity
as shown in Fig. 3. Late-type, low mass 
galaxies have a larger amount of gas (per unit mass) than early-type, 
massive discs.
 
The relationship between 
$log(M_{gas}/L_H)$ and Hubble type has a slightly smaller dispersion
if limited to the Virgo cluster galaxies which have more
reliable morphological classifications. 
The relationship with the total mass is 
clearer than with the Hubble type.
As for the birthrate parameter,
we observe a trend between $log(M_{gas}/L_H)$ and $L_H$ inside
each morphological class implying that part of  the scatter in the 
gas vs. morphology relationship is due to the large scatter
in the galactic mass within a given type. 
The residual of the $log(M_{gas}/L_H)$ vs. $L_H$
relationship, however, is not correlated to the presence of a bulge.

\subsection {The relation between gas content and the history of star 
formation in galaxies}

A relationship between the SFR and the gas density is known to
exist locally within the discs of nearby late-type galaxies (Kennicutt 1989).
Here we show that this relation extends to global quantities, integrated over 
the whole galaxy. This is not obvious a priori, since 
it is well known that most of the HI gas reservoir is located
outside the optical disc of spiral galaxies, where the star formation does 
not take place.
   
Figure 4 shows the relationship between the history of star formation 
as traced by the birthrate parameter and the total gas reservoir 
(per unit galaxy mass).

Galaxies with $b$ $\sim$ 1 have large amount of gas, while objects 
which formed most of their stars in the past ($b$ $\sim$ 0.1-0.01) 
almost exhausted 
their gas reservoirs, becoming quiescent.
Figure 4 shows a segregation between pure discs (filled circles) and bulge 
galaxies (open circles), the former being star-forming and gas-rich, 
the latter more quiescent and gas-poor.

\subsection {The SFE and the gas consumption timescale of galaxies}

The comparison of the current SFR to the gas content 
(atomic and molecular) gives an estimate of the efficiency of a given
object to transform its gas reservoir into stars. The
star formation efficiency (SFE) is defined as:

\begin{equation}
{SFE = \frac{SFR}{M_{gas}} ~~~({\rm yr^{-1}})}
\end{equation}

\noindent
No relationship is observed between the SFE and the  morphological type 
or the $H$ luminosity (Fig 5a and b). Pure exponential discs (filled circles)
have, on average, higher SFEs than bulge dominated spirals (empty circles)
of similar $H$ luminosity. If restricted to disc dominated galaxies, 
a trend is observed, with low mass galaxies having a lower SFE
than giant discs.


The SFE can vary within a given morphological class by up to
a factor 10. The average value is 4 10$^{-10}$ yrs$^{-1}$. The SFE is a time
independent quantity: it measures the $\it{instantaneous}$ efficiency of 
transforming the
gas into stars. Since the atomic gas has to go through the molecular 
phase to form
stars inside molecular clouds, a more accurate measure of the SFE can 
be obtained if
$M_{gas}$ in eq. (6) is replaced by the gas $\it{instantaneously}$ 
contributing 
to the process of star formation, thus by $M(H_2)$. Even using this 
modified
definition of the SFE, we do not observe any relationship between the 
SFE and other structural parameters, in agreement with Rownd \& Young (1999).

The SFE is proportional to the time-scale for gas depletion if 
the fraction of gas ejected by stars and recycled is taken into account.

This time-scale, generally referred to as the "Roberts' time" (Roberts 1963) 
is given by the relation (Kennicutt et al. 1994):

\begin{equation}
{\tau _R = \frac{(\frac{M_{gas}}{SFR})}{(1-R)} = \frac{(\frac{1}{SFE})}{(1-R)}}
\end{equation}

\noindent
where $R$ is the returned gas fraction. As discussed previously, the 
$R$ parameter changes with time, with the IMF and with the birthrate 
history. The determination of the future evolution of the disc is
strongly related to the dependence of $R$ on the birthrate history of
a galaxy.
Kennicutt et al. (1994) have studied how $(1-R)^{-1}$, the correction parameter
for the determination of the Roberts' time, changes for different IMFs, 
star formation laws (i.e. the relationship between the gas surface density 
and the SFR, known as the Schmidt law) and star formation efficiencies, 
these three variables being
the most important parameters for the determination of the birthrate 
history of a galaxy. Their analysis have shown that $(1-R)^{-1}$ is in the
range 1.5 - 4 for most star forming discs characterized by a birthrate 
parameter $b$=1-0.1, while it can be larger for rapidly evolving galaxies 
($b$ $<$ 0.1) with low gas surface densities. 
Since the determination of $R$ as a function of the star formation history
of a galaxy is behind the scope of the present paper, we assume a standard
correction factor $(1-R)^{-1}$ = 2.5, consistent with Kennicutt et al. (1994).

The Roberts' time 
is independent of the mass and of the morphological type (Fig. 6). This result 
is in contrast with that obtained by Donas et al. (1987), who observed a weak
trend between the Roberts' time and the morphological type,
with late-type systems showing a longer $\tau _R$. 
The time-scale for gas depletion is in the range between $\sim$ 1 Gyr and 25 
Gyr, with an average value of 10 Gyr, consistent with the value  
found by Kennicutt et al. (1994). 
These values are upper limits to the real gas consumption time scales
because a large part of the available HI gas is located outside the 
optical
disc of galaxies and thus does not contribute directly to the star formation,
unless gas inflow toward the centre of the galaxies is invoked.

\section {A model of the star formation history}

Let's summarize the evidences collected so far:

\noindent
1) photometric H$\alpha$ and UV (2000 \AA) measurements of
52 late-type (Sa-Im-BCD) galaxies, combined with
spectral evolutionary synthesis models show that the IMF of galaxies 
is consistent with a Salpeter IMF of 
slope $\alpha$ = 2.35 and an upper mass cut off of about 80 ${\rm M\odot}$, 
in the mass range $m$ $>$ 2 ${\rm M\odot}$.
Galaxies of different morphology and luminosity seem to have similar IMFs.

\noindent
2) the birthrate parameter $b$, which gives the present-to-past SFR ratio
($SFR/<SFR>_{past}$), is strongly correlated with the total mass of 
galaxies as traced by their near-IR luminosity;
the relationship between $b$ and the morphological type is weaker than with $L_H$. 

\noindent
3) the total gas reservoir per unit mass anti-correlates with the 
total mass
of galaxies; as for $b$, the relationships between $M_{gas}$/$L_H$ and 
the morphological type are more dispersed than that with $L_H$. 
Low mass, dwarf galaxies, which are generally pure exponential discs, have 
a higher gas content per unit mass and present to past SFRs than 
early-type, massive spirals.

\noindent
4) the birthrate parameter $b$ correlates with the total gas content of galaxies.

\noindent
5) the SFE and the time scale for gas depletion do not strongly correlate 
with 
properties of galaxies such as the luminosity and/or the morphological type.
If limited to pure discs, however, a trend between SFE and the total galaxy
mass is observed.
The time scale for gas depletion is $\sim$ 10 Gyrs for normal, unperturbed
late-type galaxies.

Point 1 is in agreement with the universality of the IMF in galaxies 
found in the high-mass range of normal galaxies (Scalo 1986, 1998; 
Massey 1998). Recent
results based on star counts in OB associations in the Magellanic clouds 
seem to indicate
that the IMF for massive stars is independent of metallicity 
(Hill et al. 1994).
No systematic differences have been observed in the IMFs of  
massive stars associations
in the Milky Way, LMC and SMC (Hill et al. 1994; Massey et al. 1995; 
Hunter et al. 1996c, 1997), M31 (Hunter et al. 1996a) 
and M33 (Hunter et al. 1996b); in all these nearby galaxies the IMF has 
a slope consistent with 
$\alpha$=2.5 for masses $\geq$ 1 ${\rm M\odot}$. 

Points 2 and 3 have strong implications on the models of galaxy formation 
and evolution. The importance of
the $b$ parameter resides on the fact that it gives directly an idea of the 
history of
star formation of galaxies. In galaxies with a very small value of $b$ 
(generally
early-type, massive galaxies) most of the stars
have been formed at early epochs, and the present rate of star formation 
is lower than the past one. 
This is consistent with the idea that a rapid collapse 
might have induced a
strong starburst, which efficiently transformed most of the gas 
into stars.
The lack of gas at the present epoch makes these galaxies quiescent. 
Conversely, in objects with $b$ $\sim$ 1 (late-type, dwarf, low mass galaxies) 
the gas is presently transformed into stars at the same rate than in the past.
These observational evidences are consistent with the model of galaxy 
formation discussed by Sandage (1986), who
proposed that galaxies are coeval systems, formed from the 
collapse of a primordial
gas cloud, with a collapse time-scale depending on angular momentum. 
In elliptical galaxies
the collapse was efficient enough to transform most of the gas into stars 
within a few
10$^8$ yr. In late-type systems the initial SFR was comparable with the 
present one (few solar
masses per year), so that $\sim$ 10 Gyrs after the formation of the 
primeval galaxy a large fraction
of the gas is still available to feed new stars.
This idea was recently rediscussed by Gavazzi et al. (1996a), Gavazzi 
\& Scodeggio (1996),
and Gavazzi et al. (1998). They observed that the $U-B$, $B-V$, $UV-B$ and $B-H$ colour 
indices and the H$\alpha$
equivalent width depend strongly on the galaxy mass. 
They interpreted these observational evidences 
as an indication 
that the total mass contributes to the regulation 
of the process of collapse of primeval gas 
clouds. This idea is also supported by the recent numerical simulation
by Noguchi (1999), indicating that the bulge-to-disc ratio seems 
primarily regulated by the total mass of the galaxy. A similar result 
was obtaind by Boissier \& Prantzos (2000) and Boissier et al. 
(2000), whose chemo-spectrophotometric models
of galaxy evolution reproduce several observed properties of disc galaxies 
only when the gas infall (and thus the star formation activity) is
regulated by the total mass of the galaxy.
 
In the assumption that galaxies are coeval, and that they evolved as 
closed boxes, 
Gavazzi \& Scodeggio (1996) and Gavazzi et al. (1998) were able to 
reproduce 
the relationships between the colour indices, the H$\alpha$ equivalent widths 
and the $H$ 
luminosities, using the population synthesis models of Bruzual \&
Charlot (1993) and Kennicutt et al. (1994) and adopting a star formation 
law of the type:

\begin{equation}
{SFR(t) = SFR_o e^{(-t/\tau)}    ~~~~~~~~ {\rm M\odot/yr}} 
\end{equation}

\noindent
with a Salpeter IMF,
where $t$ is the age of the galaxy, $\tau$ is the time scale for collapse,
 and $SFR_o$
is the rate of star formation at the time of the galaxy formation ($t$=0). 
They
found small values of $\tau$ ($\tau$ $\sim$ 0.5 Gyr) for giant discs and 
large 
($\tau$ $\sim$ 10 Gyr) for low mass objects.

The observational results of the present work are consistent with this 
simple picture.
Let us assume that galaxies evolve as closed boxes following a star formation 
law of the
type described in eq. (8), but assuming a more realistic age of $t_o$ = 
12 Gyrs.

$M_{star}(t)$ at a given time $t$ is given by:

\begin{equation}
{M_{star} (t)= (1-R) \int_{0}^{t} SFR_o e^{(-t'/\tau)} dt' = (1-R) SFR_o \tau [1 - e^{(-t/\tau)}]} 
\end{equation}

\noindent
\footnote{as discussed in sect. 4.1 $R$ can be taken as a constant for $t$ $>$ 1 Gyr} 
that can be directly measured from the $H$ band luminosity as described in 
sect. 4.1:

\begin{equation}
{M_{star} (t)= L_H(t) DM_{cont}(t) [M_{tot}/L_H(t)]} 
\end{equation}

\noindent
given the fact that the $H$ luminosity traces the total dynamical mass of 
galaxies.
$SFR_o$ gives the number of stars formed per year at the formation of 
the galaxy. Its dependence on the total mass implies that the process
of collapse is not scale-free. To simplify the formalism, we define 
$X(t) = DM_{cont}(t) (M_{tot}/L_H)$. As discussed in sec. 3.2, 
$X(t)$=0.5 $\times$ 4.6 at $t=t_o$=12 Gyr. 

\noindent
Combining eq. (5) with eq. (8) and (9) we can directly estimate the 
birthrate parameter $b$
as a function of $t$ and $\tau$:

\begin{equation}
{b(t)= \frac{t~e^{(-t/\tau)}}{\tau(1-e^{(-t/\tau)})}} 
\end{equation}

\noindent
The values of $b$ obtained from eq. (11) for various values of $\tau$ and
assuming $t$=$t_o$=12 Gyrs
are consistent with those obtained from the population synthesis models of 
Kennicutt et al. (1994).

\noindent
If $M_{tot}$ is the total mass of a galaxy (constant with time in a 
closed box scenario), 
we can assume that:

\begin{equation}
{M_{tot} = M_{gas}(t) + M_{star}(t) + M_{DM}(t) ~~~~~~~~\forall t}
\end{equation}

\noindent
where $M_{gas}$, $M_{star}$ and $M_{DM}$ are respectively the total gas, 
star and dark 
matter masses at a time $t$. If we assume that all the gas will be in the 
future transformed into stars, for $t$ $\rightarrow$ 
$\infty$ $M_{gas}(t)$ $\rightarrow$ 0, while $M_{star}(t)$ = 0 for $t$=0,
thus $M_{star}({\infty})$ = $M_{gas}(0)$ = $M_{tot}$ - $M_{DM}(0)$ = 
$SFR_o$$\tau$ (assuming that all the returned gas is used to form stars).
If we assume that $M_{DM}$ does not change with time, $M_{gas}(t)$ 
can be determined from relation (12).

\noindent
By substituting $SFR_o$$\tau$ to $M_{tot}$ - $M_{DM}$, eq. (12) becomes:

\begin{equation}
{M_{gas}(t) = SFR_o \tau e^{(-t/\tau)}}
\end{equation}

\noindent
and

\begin{equation}
{M_{gas}(t)/L_H(t) = \frac {X(t)}{(1-R)}\frac{e^{(-t/\tau)}}{(1 - e^{(-t/\tau)})}}
\end{equation}

\noindent
in the assumption that $R$ does not change with time.

At the same time this simple model predicts that the SFE defined in eq. 
(6) should be:

\begin{equation}
{SFE = \tau^{-1}}
\end{equation}

\noindent
thus independent on $t$. This means that each galaxy should have had an 
efficiency in transforming gas into stars which depends only on the $\tau$
parameter, but similar at any epoch.

This analytic representation of all the observed variables used in this
work ($SFR$, $b$, $M_{gas}$, $L_H$, $SFE$) can be done once a star formation 
law such as that given in eq. (8) is assumed
\footnote{We have repeated this exercize replacing the exponential 
star formation law (eq. 8) with a ``delayed exponential'' law, such as the
one proposed by Sandage (1986). This law better accounts for cases with
increasing SFR with time, as observed in some irregular galaxies
such as IZw18 (Kunth \& Ostlin 2000). We found results qualitatively consistent
with the exponential case, thus we avoid to expand the ``delayed  
exponential'' case in full details}.

 
The analytical model can be compared to the observables only once $SFR_o$
is known, $L_H$ being a function of $SFR_o$; the other variables 
($b$, $M_{gas}$/$L_H$, $SFE$), being normalized entities, do not depend on
$SFR_o$.
This parameter, which gives the rate of star formation  
at the time of the galaxy formation ($t$=0), is clearly dependent 
on the mass of the galaxy ($SFR_o$ $\times$ $\tau$ = $M_{star}$ + $M_{gas}$).

Following Gavazzi \& Scodeggio (1996), if we assume a simple empirical 
relationship between the $H$ luminosity and the exponential decay time 
scale for star formation of the type:

\begin{equation}
{log L_H = a \times log \tau + c}  
\end{equation}

\noindent 
we can derive from eq. (9) an (10) a semi-analytical relationship between 
$SFR_o$ and $\tau$:

\begin{equation}
{SFR_o = \frac {X(t_o) \tau^a 10^c}{(1-R) \tau e^{(-12/\tau)}}}
\end{equation}

\noindent
Once $SFR_o$ is determined, we can compare the
predictions of the analytical models with 
the relationships previously discussed
between $b$, $M_{gas}$/$L_H$ or $SFE$ and the $H$ band luminosity 
(Fig 7 a,b,d). The model prediction (dotted line) 
can be directly 
compared to the observations for the $M_{gas}$/$L_H$ vs. $b$ relationship 
(Fig. 7 c), without any assumption on the 
relationship between $\tau$ and $L_H$,
both being normalized entities thus independent of $SFR_o$.
At the same time the obvious observed scaling relationships between 
$SFR$, $M_{gas}$ and $L_H$ can be compared to the models prediction 
(Fig. 7 e,f) and used, together with the normalized entities, 
to constrain the free parameters $a$ and $c$.
An excellent fit to the data is obtained for $a$ = -2.5 and $c$ = 12 once 
$\tau$ is expressed in Gyrs 
consistent with  Gavazzi \& Scodeggio 
(1996; $a$ = -2.5; $c$=11.12).

In conclusion, the observed relationships between optical, near-IR, 
UV colours, 
H$\alpha$ E.W.s, the birthrate
parameter $b$, the total gas mass and the $H$ luminosity can, at least 
to the first order, be reproduced in the simple assumption
that galaxies are coeval ($\sim$ 12 Gyrs), that they evolved as 
closed boxes following an exponentially declining 
star formation law, with a decay time scale 
depending on the
mass of the primeval cloud, small ($\tau$ $\sim$ 0.5 Gyrs) for massive 
objects ($L_H$ = 10$^{12}$
$L_{H\odot}$), and large ($\tau$ $\sim$ 10 Gyrs) for dwarf systems ($L_H$ 
= 10$^{9}$$L_{H\odot}$).

If the models here discussed are realistic, we conclude that the strong 
relationship
observed between various star formation tracers, such as the UV flux
(Donas et al. 1990; Deharveng 
et al. 1994; Buat 1992; Boselli 1994), the H$\alpha$ flux (Kennicutt 1989; 
1998; 
Scodeggio \& Gavazzi 1993 and references therein) and the HI or total gas 
surface density in 
unresolved galaxies is not due to a local relationship between the 
gas column density
and the process of star formation, known as the Schmidt law, but it is simply 
a consequence of the closed-box evolution. 
From eq. (8) and (13) we expect in fact that the integrated SFR 
and the total gas content decrease with time with a similar  
decline. This conclusion 
is corroborated by the fact that in "normal", unperturbed galaxies the HI gas 
(which  constitutes
$\sim$ 90\% of the total gas reservoir; Boselli et al. 1997b) is 
found in a flat disc of diameter about twice as large as the optical one 
(Cayatte et al. 1990). This gas
is not expected to contribute directly to the star 
formation since 
i) up to 75\% of the HI reservoir is located outside the optical 
disc where the star formation takes place; ii) the HI has to go 
through the molecular phase before forming stars.
Thus an infall of gas from the outer HI disc to the optical
disc must be present.
Spiral galaxies of all morphological types and luminosity still have 
enough gas to continue their present day star formation
activity for $\sim$ 10 Gyrs. This time is significantely longer than previous
estimates because we assume an higher contribution of the recycled gas 
(see Kennicutt et al. 1994) and because the HI-deficient
cluster galaxies are excluded from the analysis (Boselli 1998). 





\section{Comparison with observations at high redshift}

The aim of this section is to discuss if the closed-box scenario
discussed in the present paper, based on observations of local "adult"
late-type galaxies, is consistent with observations at larger redshift.

Several recent attempts to reconstruct the evolution
with look-back time of the total star formation activity 
of the Universe (Cowie et al. 1999; Steidel et al. 1999; Madau et al. 1998
and references therein) unanimously concluded that
it increases by about an 
order of magnitude from $z=0$ to $z$ $\sim$ 1. 
The dependence for $z>1$ is more controversial, however the most
recent determination by Steidel and collaborators (1999) indicates that, 
if the data are
appropriately corrected for extinction, the star formation activity 
of the Universe
remains roughly constant for larger look-back times, up to $z\sim 4$.

Eq. (8), together with eq. (16) give the time evolution of the $SFR$ of 
galaxies of luminosity $L_H$, for a given $SFR_o$. 

We can predict, for a galaxy of a 
given $H$ luminosity at $z$=0, its $SFR$ at any $z$. 
We can then estimate the SFR per unit covolume of the Universe by integrating 
the contribution to
the SFR of each galaxy, weighted according to a given luminosity 
function.
Since we are dealing with late-type galaxies, 
we assume the local $B$ band Schechter luminosity function of Heyl et al. (1997) 
for types Sa-Sm
and transform $B$ magnitudes into $H$ magnitudes assuming the 
$B-H$ vs. $B$ colour-magnitude relation: 
 
\begin{equation}
{B-H = 7.7 -0.38 \times B}
\end{equation}

\noindent
as determined from the subsample of Virgo late-type galaxies which spans 
the full dynamic range in luminosity.

The resulting SFR per comoving volume of the Universe 
is compared with the observed SFR as a function of $z$ in Fig. 8. 
For consistency with previous works we assume $\Lambda$=0,
$H_o$=50 km s$^{-1}$ Mpc$^{-1}$ and $q_o$=0.5.
The ``observed'' values of $SFR$ (in ${\rm M\odot yr^{-1} Mpc^{-3}}$)
from Steidel et al. (1999) (open symbols)
are corrected for extinction. 
The values of the 2800~ \AA~
luminosity density given by Cowie et al. (1999) (filled squares)
are transformed into ${\rm M\odot yr^{-1} Mpc^{-3}}$ 
and corrected for extinction consistently with Steidel et al. (1999). 
The value of Treyer et al. (1998) (filled triangle) 
and Gallego et al. (1996; filled circle) obtained respectively from the 
local UV at 2000 \AA ~ and H$\alpha$ luminosity functions, are also
transformed into ${\rm M\odot yr^{-1} Mpc^{-3}}$ as described in
sect. 3, adopting for simplicity an extinction correction of 0.3 mag. 
Such a small UV
and H$\alpha$ extinction derives from the assumption that the UV and H$\alpha$
luminosity functions are dominated by low mass, low 
luminosity, blue galaxies (see Appendix). 

Figure 8 shows that, at least qualitatively, the model predictions 
give an excellent fit to the updated 
estimates of the SFR per comoving volume.

Our model (dotted line) 
is in excellent agreement with the points at $z$=0 
by Gallego et al. (1996; filled circle) and by Treyer et al.(1998). 
It accurately reproduces the relation $SFR~Mpc^{-3}$ $\alpha$ (1+$z$)$^{1.5}$
observed by Cowie et al. (1999) in the range
0 $<$ $z$ $<$ 1.5 (filled squares), and is consistent with 
the extinction-corrected estimates of Steidel et al. (1999) for 
3 $<$ $z$ $<$ 4.5. 

It is remarkable that the predicted SFRs are in agreement with the 
observations, even though we consider only the contribution
of the late-type galaxies to the evolution of the SFR of the Universe. 
One should remember however that the adopted SFR per comoving volume, 
with which we compare our model predictions,
have all been estimated from UV-selected samples, thus
biased towards low-extinction galaxies. Elliptical galaxies, which are
expected to be formed in violent starbursts with strong dust extinctions, 
should not be present in UV-selected samples.
Their contribution at early epochs should be however observable in the 
millimetric domain (Franceschini et al. 1994).

Using a similar procedure we can compare the model predictions with the 
observed comoving density of the gas,
as determined from observations of damped Ly$\alpha$ systems. As above,
the contribution to the total gas density of the Universe is weighted 
according to the $B$ luminosity
function of Heyl et al. (1997), and $B$ magnitudes are transformed into $H$
magnitudes using eq. (18).  
The $\tau$ parameter in eq. (13) 
is constrained
by the relationship between $\tau$ and the $H$ luminosity given in eq. (16) 
(with $a$=-2.5 and $c$=12).
The gas density of the Universe at different epochs
can be estimated by integrating the derived "gas mass" luminosity function.

The models predictions are found in remarkable agreement (see Fig. 9) 
with the gas density of the
Universe $\Omega$$_g$ (expressed in units of the present critical density) 
as determined from the statistical analysis of 
damped Ly$\alpha$ systems by Pei et al. (1999) and Rao \& Turnshek (2000).
We use the values of $\Omega$$_g$ given by Pei et al. (1999) 
corrected to take into account the missing damped Ly$\alpha$ 
systems for dust obscuration in an optically selected sample of quasars.
To be consistent with Pei et al. (1999) we removed the contribution
of H$_2$ (10\%) to the values of $\Omega$$_g$ estimated from our model.   
Both models predict a gas density of the Universe at $z$=0 $\sim$ 4 times
higher than the observed value determined from the local HI luminosity 
function of Zwaan et al. (1997); this is due to the fact that the models 
overestimate the gas content of low luminosity galaxies (see Fig. 7f),
whose contribution to the local luminosity function is dominant.

The agreement between $\Omega$$_g$ predicted by our model and the
result of the statistical analysis of damped Lyman alpha systems is good
at redshifts $<$ 2, while it becomes poor at higher $z$.
The determination of $\Omega$$_g$ at $z$ $>$ 1.5 strongly depends 
on the adopted correction for missing absorbers at high redshifts, and thus,
as for the SFR, on the variation of the dust to gas ratio with $z$. 
The disagreement between the model prediction and $\Omega$$_g$ 
for $z$ $>$ 2 could thus be due to an underestimate of the
extinction bias at high redshifts in the optically selected sample 
of quasars \footnote{
We notice that the extinction correction assumed by Pei et al. (1999)
to correct $\Omega$$_g$ for missing absorbers is lower than that 
used by Steidel et al. (1999) to estimate the SFR at high $z$.
}. \\
Pei et al. (1999) interpret the decrease of $\Omega$$_g$ 
with $z$ as due to the fact that damped Lyman alpha systems are galaxies
which grow by accreation of (essentially ionized) gas. This would result 
into a decrease of the co-moving neutral gas density of the Universe with
redshift. Our closed-box model cannot reproduce any decrease of
$\Omega$$_g$ with $z$ since the gas is always assumed in the neutral phase.

We remark that the model predictions at high redshift
depend strongly on the parameters $a$ and $c$ of eq.(16) and on the 
assumed $B$ luminosity function.
Furthermore the absolute value of the $SFR~Mpc^{-3}$ at various $z$
might change by a factor of $\sim$ 2 when adopting different population 
synthesis models.

\medskip

To summarize, this work has shown the importance of mass in
parametrizing most of the physical properties of nearby galaxies.
The history of star formation and the amount of total gas (per unit mass)
of galaxies are found to depend primarily on their total mass and only 
secondly on their morphological type. These strong observational evidences
must be reproduced by the models of galaxy formation and evolution.
 
The analysis carried out in the present paper can be
considered as an exercize aimed at showing that the "monolithic" scenario
of galaxy formation is not yet observationally falsified, both on
local and on cosmological scales.
Several scaling relations observed in nearby galaxies (i.e.
the gas content, the activity of star formation and the stellar populations)
can be reproduced assuming that galaxies are coeval and that they 
evolved as closed boxes with an universal star formation law
whose characteristic time scale parameter $\tau$ 
scales inversely with the total mass of the parent galaxies.
The same scenario predicts a look-back time dependence of the 
integrated star formation activity and gas content of the Universe 
consistent with the observations up to large redshifts. 
In order to reproduce the observed local properties of galaxies
and the cosmological dependences discussed above, we see no compelling 
need for more sophisticated galaxy formation models, such as the 
hierachical model proposed by Kauffmann et al. (1993) and by Cole et al. 
(1994). 

The present work seems to indicate that the evolution of the star 
formation rate and the gas density of the Universe is due to
the gas consumption via star formation processes according to
a simple, universal star formation law, independent on $z$, valid
for all rotating systems. 

Beside any cosmological speculation, the
firm conclusion of the present work is that the history of star formation 
of disc galaxies is primarily regulated by their total mass and only 
secondly by their angular momentum.

\begin{acknowledgements}

We want to thank J. Lequeux for his precious comments and suggestions
which helped improving the quality of this paper.
We wish to thank JM. Deharveng, V. Buat, G. Comte, JM. Deltorn, B. Milliard
 and M. Treyer for interesting discussions, S. Charlot
for making some unpublished parameters derived from his stellar 
population synthesis models available to us. \\
This work is based on observations taken at the Observatorio Astron\'omico 
Nacional (OAN), San Pedro M\'artir, B.C. (M\'exico), at
the Observatoire de Haute Provence (OHP) (France), at the TIRGO (Gornergrat, 
Switzerland), at the Calar Alto Observatory (Spain) and at the 12-m National 
Radio Astronomical Observatory (NRAO), Kitt Peak (Arizona). 
The OAN is operated by the Universidad Nacional Aut\'onoma de M\'exico (UNAM),
the OHP is operated by the CNRS, France,
TIRGO is operated by CAISMI-CNR, Arcetri, Firenze, Italy.
Calar Alto is operated 
by the Max-Planck-Institut f\"ur Astronomie (Heidelberg) jointly with the 
Spanish National Commission for Astronomy.
The NRAO is a facility of the National Science Foundation operated
under cooperative agreement by Associated Universities, Inc.

\end{acknowledgements}

\appendix

\section{Corrections to H$\alpha$+[NII] and UV fluxes}

\subsection{The [NII] contamination}

The analysis of the integrated spectra of late-type 
galaxies by Kennicutt (1992)
has shown that, on average, the $[NII](6583 \AA)/H\alpha$ 
emission line ratio is $\sim$ 0.5. The sample
of Kennicutt (1992), however, is dominated by active galaxies 
(see sect. A.2).
In order to estimate a $[NII]/H\alpha$ ratio representative of 
normal galaxies, as those 
analysed in the present work, we plot in Fig. 10 the $[NII]/H\alpha$ 
ratio for
spirals and irregular galaxies in the sample of Kennicutt (1992) 
as a function of the Hubble type, 
excluding peculiar objects such as Markarian galaxies.

The figure clearly shows a different trend between spirals and irregulars, 
with an average $[NII]/H\alpha$ = 0.42 $\pm$ 0.19 for Sa$\le$ type $\le$ 
Scd and
$[NII]/H\alpha$ = 0.25 $\pm$ 0.15 for $\ge$ Sd. These values are 
consistent
with the result of Kennicutt (1983) based on high resolution spectroscopy
of bright HII regions in nearby galaxies:  
$[NII]/H\alpha$ = 0.33 $\pm$ 0.12 for spirals and
$[NII]/H\alpha$ = 0.08 $\pm$ 0.05 for irregulars.
In the present analysis we correct the H$\alpha$ fluxes for [NII] 
contamination
using $[NII]/H\alpha$ = 0.42 for galaxies of type $\le$Scd (including 
Peculiar galaxies and unclassified spirals) and 
$[NII]/H\alpha$ = 0.25 for galaxies of type $\ge$ Sd.

\subsection{The UV and H$\alpha$ extinction in normal galaxies}

H$\alpha$ and UV fluxes can be used to estimate the rate of star formation 
in galaxies
provided that they are corrected for internal extinction. 
Independent works based on different techniques indicate that 
the average extinction of 
the H$\alpha$ and UV (2000 \AA) fluxes of normal galaxies are similar, 
despite 
a difference by a factor $\sim$ 4 in the extinction coefficient 
$k_{\lambda}$,
as determined from the galactic extinction law 
(Savage \& Mathis 1979;
Bouchet et al. 1985). This observational evidence is generally 
explained as a 
result of a combined effect of a standard extinction curve with
a different geometrical distribution of emitting stars and absorbing 
dust at 2000 \AA~ and in H$\alpha$. 
While the H$\alpha$ emission of late-type galaxies is dominated
by the photoionization of gas by hard UV photons produced by massive 
($m$ $\ge$10 ${\rm M\odot}$) O-B stars (Kennicutt 1998) located 
inside HII regions, 
the UV emission at 2000 \AA~ is dominated by A stars of intermediate 
mass (2 $\le$ $m$ 
$\le$ 5 ${\rm M\odot}$) (see sect. 3), old enough to have migrated outside the HII 
regions. This has been observed for example in
M33 (Buat et al. 1994), where $\sim$ 20\% of the total UV emission at 
2000 \AA ~ is diffuse, even though part of the diffuse emission
might be scattered light. Given the complex distribution of 
absorbing dust and emitting stars, the determination of the
fraction of scattered light in spiral galaxies is very uncertain;
however the models of Witt \& Gordon (2000) indicate that the scattered light
in the UV at 2000 \AA ~is $\sim$ 20\% once a clumpy geometry is assumed.  
Furthermore the observations of nearby galaxies are limited to few 
late-type spirals, preventing us to 
quantify the contribution of the diffuse emission to the 
total UV emission in early-type spirals. It is however conceivable that 
in Sa the diffuse emission contribution is 
even higher than in Scd galaxies such as M33. 
Since dust is mostly associated with HII regions, 
the relative geometrical
distribution of absorbing dust and emitting stars, which increases the 
absorption of
H$\alpha$ with respect to UV photons at 2000 \AA, should compensate 
the $\sim$ 4 times higher UV extinction with respect to extinction at 
H$\alpha$. 

From the analysis of the Balmer decrement
of a large sample of galaxies with long-slit integrated spectroscopy, 
Kennicutt (1992) estimates that the average extinction of the H$\alpha$ 
flux of late-type
galaxies is $\sim$ 1 mag. Consistently the comparison of free-free 
radio fluxes and H$\alpha$
fluxes of normal galaxies by Kennicutt (1983) indicates an  
extinction of 1.1 mag. 
In order to study whether the H$\alpha$ extinction is
morphological-type dependent, we have re-examined the sample of Kennicutt 
(1992)
limited to "normal" galaxies (thus excluding Markarian objects), 
dividing 
galaxies in two subsamples: Sa$\le$ type $\le$ Scd and type $\ge$ Sd. 
High resolution 
spectra were selected as first priority. The extinction of the H$\alpha$ 
flux 
strongly depends on the assumption on the underlying Balmer absorption, 
in particular in 
normal galaxies where the H$\beta$ line is generally weak. Values 
found in the literature range between 2 and 5 \AA~ in $H\beta E.W.$ 
(Kennicutt 1992; McCall et al. 1985).
We calculate for simplicity
the H$\alpha$ extinction using these two values of the underlying 
Balmer absorption. Including this correction and
following the prescription of Lequeux et al. (1981) we can estimate that the 
extinction at a given
$\lambda$ from the Balmer decrement is:

\begin{equation}
{A({\lambda}) = 2.5 C({\lambda}) }
\end{equation}

\noindent
where

\begin{equation}
C({\lambda}) = \frac{1}{f(H{\alpha})}\biggl[log\frac{I(H{\alpha})}
{I(H{\beta})} - log\frac{F(H{\alpha})}{F(H{\beta})} + 
log\bigl(1+ \frac{E}{H{\beta}E.W.}\bigr)\biggr]
\end{equation}

\noindent
where $f(H{\alpha})$ is the galactic reddening function normalized at 
H$\beta$, $f(H{\alpha})$=-0.355
(Lequeux et al. 1981), $I(H{\alpha})$ and $I(H{\beta})$ are the extinction 
corrected intensities
and $F(H{\alpha})$ and $F(H{\beta})$ are the observed intensities of the 
H$\alpha$ and H$\beta$ lines.
$E$ is the underlying Balmer absorption (in \AA). Assuming a ratio of 
$I(H{\alpha})$/$I(H{\beta})$
=2.86 for a case B nebula at 10$^4$ K (Brocklehust 1971), we obtain an 
average value of
the H$\alpha$ extinction of $A(H{\alpha})$ = 0.78 $\pm$ 0.47 
(for $E$ = 5 \AA) and 
$A(H{\alpha})$ = 1.60 $\pm$ 0.77 (for $E$ = 2 \AA) for spirals, and
$A(H{\alpha})$ = 0.41 $\pm$ 0.29 (for $E$ = 5 \AA) and $A(H{\alpha})$ = 
0.86 $\pm$ 0.42 (for $E$ = 2 \AA) for galaxies with type $\ge$ Sd 
(see Fig. 11).

We did not observe any relation of $A(H{\alpha})$ 
or H$\alpha$ surface brightness (extended to all
the galaxies of our sample) with inclination, even once galaxies
are divided into different morphological class bins.
We stress however the fact that the sample of Kennicutt (1992) is 
biased 
towards strongly star forming galaxies, even if Markarian galaxies 
are excluded, and thus it could be not
representative of the sample of galaxies analysed in this work. 
In fact the average
$H\alpha+[NII]E.W.$ in the sample of Kennicutt (1992) is 82 \AA~ 
for spirals and 105 \AA~
for types $\ge$ Sd, while in the sample analysed 
in this work the average values are
$H\alpha+[NII]E.W.$ = 22 \AA~ and $H\alpha+[NII]E.W.$ = 46 \AA~ 
respectively. In conclusion
we decided to assume a standard H$\alpha$ extinction correction 
of 1.1 mag for
Sa$\le$ type $\le$ Scd, and 0.6 mag for type $\ge$ Sd.

The lack of any relationship between the UV surface brightness and
inclination observed for our sample galaxies prevents us to estimate 
an inclination-dependent extinction correction law. 
As shown by Buat \& Xu (1996), the UV extinction at 2000 \AA~ is 
$\sim$ 0.9 mag for galaxies with type $\le$ Scd and $\sim$ 0.2 for 
galaxies with type $\geq$ Sd. If however the ratio between the H$\alpha$ 
(corrected for [NII] contamination) 
and the UV fluxes for galaxies with type Sa-Scd is plotted versus
their axial ratio, a residual trend is still marginally present (see Fig. 12).
 
A linear fit to this relation gives:

\begin{equation}
{log (L_{H{\alpha}}/L_{UV}) = - 0.56\times log (b/a) + 1.24}
\end{equation}

\noindent
where $log (b/a$) is the axial ratio. If we make the reasonable 
assumption that 
this trend is due to increasing extinction with the inclination,
this relation can be used to determine an empirical correction to 
the UV data as a function 
of the axial ratio. This correction, however, represents a lower limit
since it has been determined assuming that the H$\alpha$ extinction
is independent of the galaxy inclination.
For a random distribution of ($b/a$) we derive an average UV extinction of
$A(UV)_i$ $\sim$ 0.3 mag. In order to be consistent with Buat \& Xu 
(1996), 
we assume:

\begin{equation}
{A(UV) = A(UV)_{f.o}(ty) - k_i(ty) \times log(b/a)}
\end{equation}

\noindent
where the values of $A(UV)_{f.o}$ and $k_{i}$ are given in Table 5 for 
different morphological classes.
Given the systematic difference in luminosity between spirals and dwarfs,
the adopted corrections takes into account, at least to the first
order, a possible dependence between extinction and luminosity.


\clearpage



\figcaption
{The logarithm of the ratio of the H$\alpha$ to UV (2000 \AA) ~fluxes versus
a) the morphological type, b) the $H$ luminosity. To avoid overplotting, 
a random number between -0.4 and 0.4 has been added to each numerical type.
Filled circles are for disc dominated galaxies ($C_{31}$ $<$ 3), open 
circles for bulge dominated galaxies ($C_{31}$ $>$ 3). H$\alpha$ and UV fluxes
are corrected for extinction and [NII] contamination.
\label{fig1}}
 
\figcaption
{The relationship between the logarithm of the birthrate parameter $b$ and 
a) the morphological type, b) the $H$ luminosity. Symbols
as in Fig 1. \label{fig2}}

\figcaption
{The relationship between the logarithm of the total gas mass 
(normalized to the $H$ luminosity) and 
a) the morphological type, b) the $H$ luminosity. Symbols
as in Fig 1.
\label{fig3}}

\figcaption
{The relationship between the logarithm of the birthrate parameter $b$ and 
the total gas mass normalized to the $H$ luminosity (in solar units). Symbols
as in Fig 1. 
\label{fig4}}

\figcaption
{The relationship between the logarithm of the $SFE$ and 
a) the morphological type, b) the $H$ luminosity. Symbols
as in Fig 1.
\label{fig5}}
 
\figcaption
{The distribution of the time for gas depletion (Roberts' time) in 6 
classes of morphological type. \label{fig6}}
 
\figcaption
{The comparison between the model prediction 
and the observations for the a) $b$ vs. $log L_H$, b) $log M_{gas}/L_H$ vs. 
$log L_H$, c) $b$ vs. $log M_{gas}/L_H$, d) $SFE$ vs. $log L_H$,
e) $SFR$ vs. $log L_H$ and f) $M_{gas}$ vs. $log L_H$ relationships. 
Symbols as in Fig 1.
\label{fig7}}

\figcaption
{The relationship between the SFR per comoving volume of the Universe 
and $z$ compared to the model predictions (dotted line); 
the extinction corrected values are from:
Lilly et al. (1996) [open circles], Connolly et al. (1997) [open squares], 
Madau et al. (1997) [open triangle], Steidel et al. (1999) [open stars], 
Treyer et al. (1998) [filled triangle], Cowie et al. (1999) [filled square], 
Gallego et al. (1996) [filled circle]. 
\label{fig8}}
 
\figcaption
{The relationship between the gas density per comoving volume of the Universe 
and $z$ compared to the model predictions (dotted line). 
The gas densities observed from damped Ly$\alpha$ systems,
corrected for missing absorbers, are from
Pei et al. (1999; filled triangles) and  Rao \& Turnshek (2000; open 
circles).
\label{fig9}}

\figcaption
{The relationship between the $[NII]/H\alpha$ ratio and the Hubble type 
for 
the "quiescent" galaxies from the sample of Kennicutt (1992). The large 
symbols represent averages. \label{fig10}}

\figcaption
{The relationship between the H$\alpha$ extinction as determined from 
the Balmer decrement using an underlying Balmer absorption of 
$H\beta E.W.$ = 2 \AA ~(top) 
and $H\beta E.W.$ = 5 \AA ~(bottom) and the Hubble type for 
the "quiescent" galaxies in the sample of Kennicutt (1992). 
The large symbols represent averages. \label{fig11}}
 
\figcaption
{The relationship between the H$\alpha$ to UV flux ratio and the 
axial ratio. The 
dashed line shows the best fit to the data. \label{fig12}}






\clearpage

\begin{deluxetable}{llccccc}
\footnotesize
\tablecaption{The Virgo Sample\tablenotemark{a}}
\tablewidth{0pt}
\tablehead{
\colhead{mag. limit} & \colhead{No. objects}   & \colhead{Halpha}   &
\colhead{UV} &
\colhead{CO}  & \colhead{HI} & \colhead{$H$ or $K$} 
}
\startdata
11         &     3      &     3  &  3 &  3 &  3 &  3       \\  
12         &     8      &     8  &  6 &  8 &  8 &  8       \\  
13         &    25      &    22  & 19 & 19 & 25 & 25       \\  
14         &    45      &    29  & 27 & 23 & 45 & 42       \\  
15\tablenotemark{b}& 54 &    32  & 33 & 23 & 54 & 51       \\  
16\tablenotemark{b}& 59 &    35  & 34 & 23 & 59 & 56       \\  
\enddata
\tablenotetext{a}{including only objects with HI-deficiency $\leq$0.3}
\tablenotetext{b}{limited to the "ISO" sample}
\end{deluxetable}
 
\clearpage 

\begin{deluxetable}{llccccc}
\footnotesize
\tablecaption{The Coma/A1367 supercluster and Cancer sample\tablenotemark{a}}
\tablewidth{0pt}
\tablehead{
\colhead{mag. limit} & \colhead{No. objects}   & \colhead{Halpha}   &
\colhead{UV} &
\colhead{CO}  & \colhead{HI} & \colhead{$H$ or $K$} 
}
\startdata
14         &      7      &     7  &  - &  7 &  7 &  7       \\  
14.5       &     20      &    20  &  5 & 17 & 20 & 20       \\  
15         &     63      &    50  & 13 & 42 & 63 & 63       \\  
15.7       &    174      &   117  & 34 & 66 &174 &174       \\  
\enddata
\tablenotetext{a}{including only objects with HI-deficiency $\leq$0.3}
\end{deluxetable}

\clearpage

\begin{deluxetable}{llcc}
\tablecaption{The statistics in different morphological bins.}
\tablewidth{0pt}
\tablehead{
\colhead{type}   & \colhead{No. objects}   &
\colhead{No. H$\alpha$ or UV}   
}
\startdata
 Sa                &   26  & 18 \\
 Sab               &   14  & 13 \\
 Sb                &   29  & 19 \\
 Sbc               &   36  & 19 \\
 Sc                &   66  & 50 \\
 Scd               &    5  &  3 \\
 Sd                &    3  &  2 \\
 Sdm - Sd/Sm       &    2  &  2 \\ 
 Sm                &    4  &  3 \\
 Im - Im/S         &    3  &  2 \\
 Pec               &   26  & 21 \\
 S/BCD - dS/BCD    &    0  &  0 \\
 Sm/BCD            &    2  &  2 \\
 Im/BCD            &    1  &  0 \\
 BCD               &    2  &  1 \\
 S (dS)...         &   12  &  5 \\
 dIm/dE            &    1  &  0 \\
 ? 	           &    1  &  1 \\ 
\enddata
\end{deluxetable}

\clearpage

\begin{deluxetable}{lcllc}
\tablecaption{Adopted IMF parameters}
\tablewidth{0pt}
\tablehead{
\colhead{IMF slope} &\colhead{IMF cut-off} & \colhead{$K_{H\alpha}(\alpha,M_{up})$\tablenotemark{a}} & \colhead{$K_{UV}(\alpha,M_{up})$\tablenotemark{b}} & \colhead{$log  [K_{H\alpha}(\alpha,M_{up})/K_{UV}(\alpha,M_{up})]$}   
}
\startdata
1.5	&	80		&	1/1.61$\times$10$^{42}$ & 1/2.01$\times$10$^{40}$& 1.903 \\
2.5	&	40		&	1/5.41$\times$10$^{40}$ & 1/3.18$\times$10$^{39}$& 1.231 \\
2.5	&	80		&	1/1.16$\times$10$^{41}$ & 1/3.54$\times$10$^{39}$& 1.514 \\
2.5	&	120		&	1/1.60$\times$10$^{41}$ & 1/3.66$\times$10$^{39}$& 1.640 \\
3.5	&	80		&	1/5.53$\times$10$^{38}$ & 1/1.67$\times$10$^{38}$& 0.520 \\
\enddata
\tablenotetext{a}{from Charlot \& Fall (1993), for $M_{low}$ = 0.1 ${\rm M\odot}$, 
in units of $({\rm M\odot yr^{-1})/(erg s^{-1})}$}
\tablenotetext{b}{from Charlot, private communication, 
for $M_{low}$ = 0.1 ${\rm M\odot}$, in units of 
$({\rm M\odot yr^{-1})/(erg s^{-1}\AA^{-1})}$
}
\end{deluxetable}

\clearpage

\begin{deluxetable}{lccc}
\tablecaption{Adopted H$\alpha$ and UV extinction parameters}
\tablewidth{0pt}
\tablehead{
\colhead{Type} & \colhead{$A(H\alpha)_{f.o.}$} & \colhead{$A(UV)_{f.o.}$} & \colhead{$k_{i}(ty)$}
}
\startdata
Sa-Scd (Pec, S... at D$>$30 Mpc)& 1.1  & 0.60 & 0.56 \\
Sd-Im-BCD			& 0.6  & 0.20 & 0.00 \\
\enddata
\end{deluxetable}






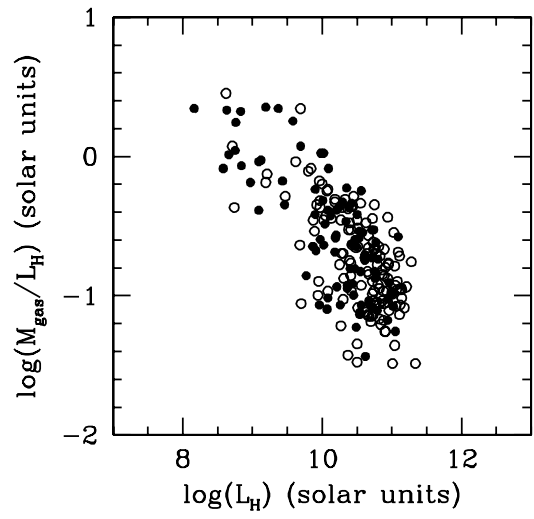
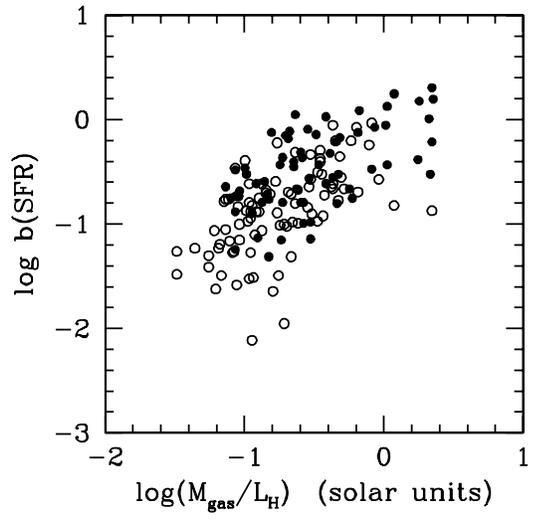
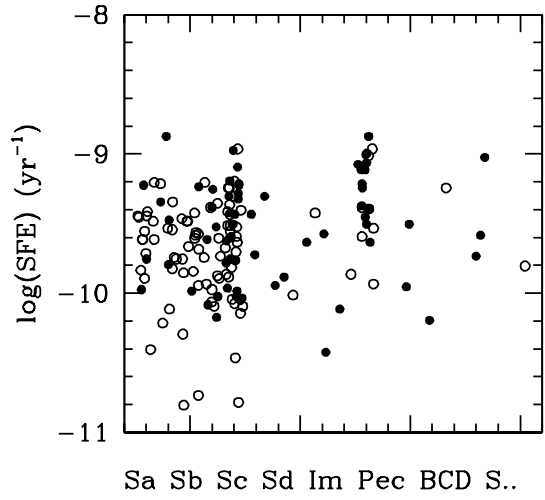
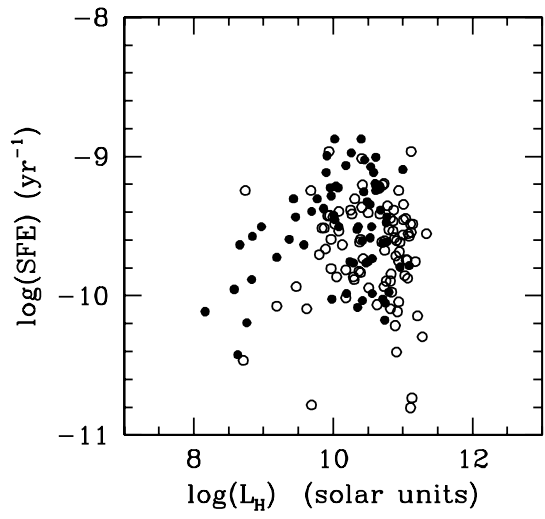
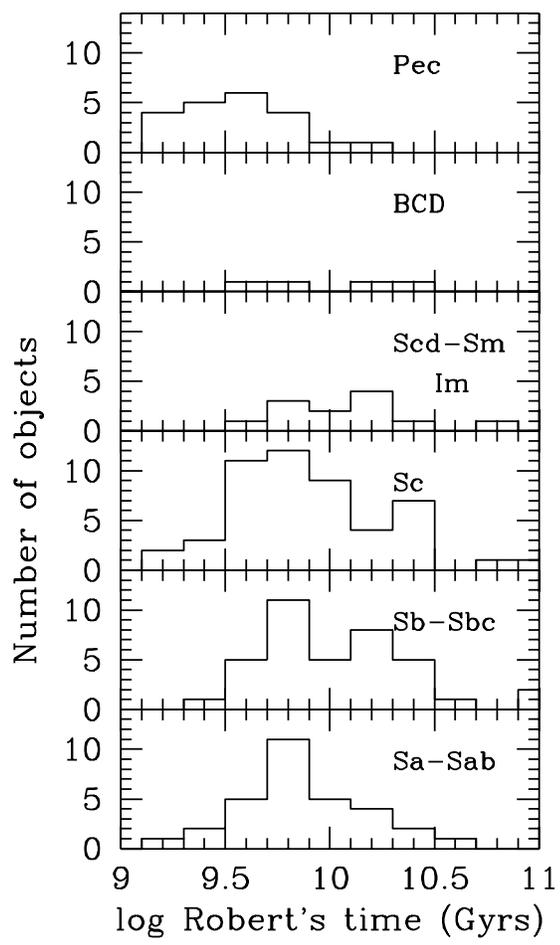
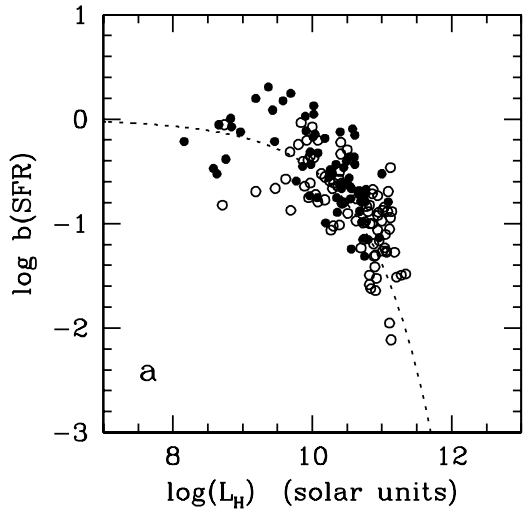
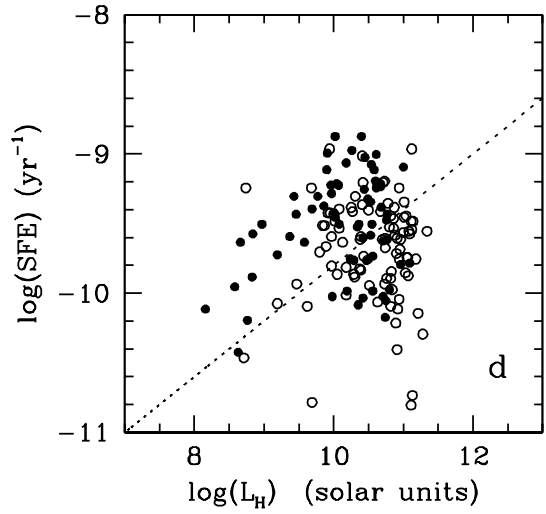
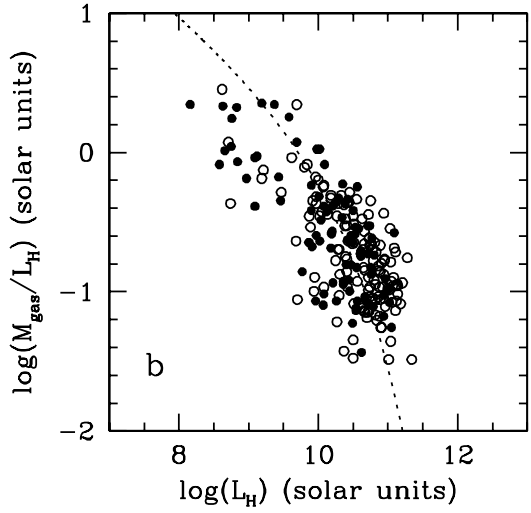
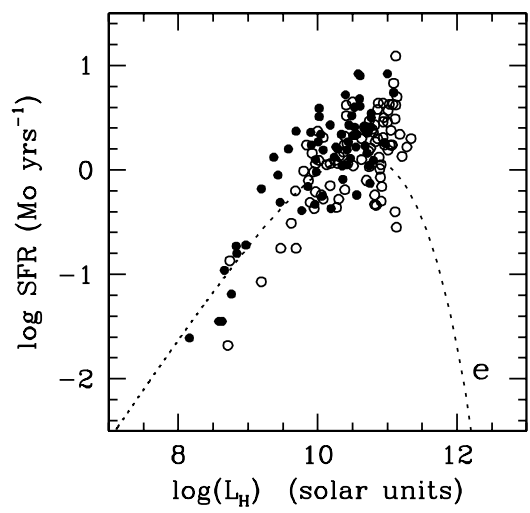
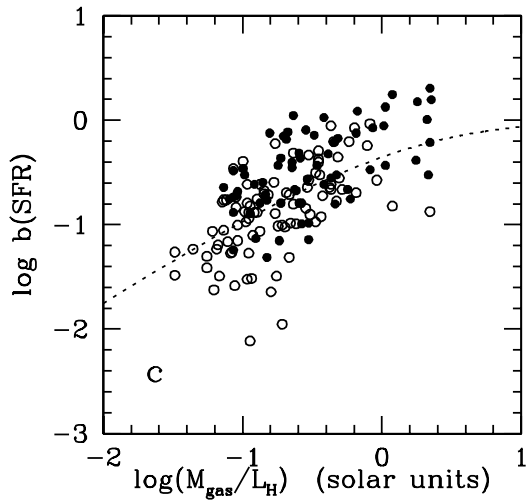
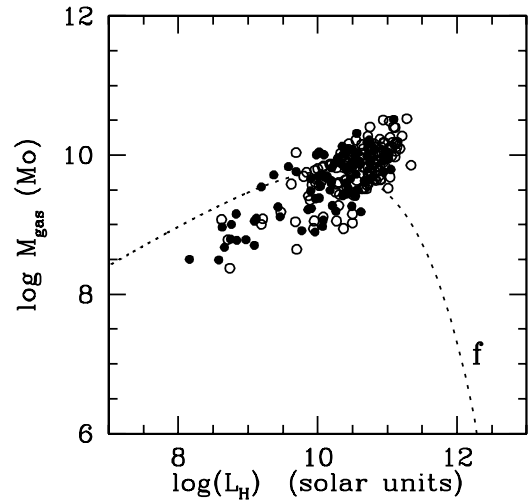
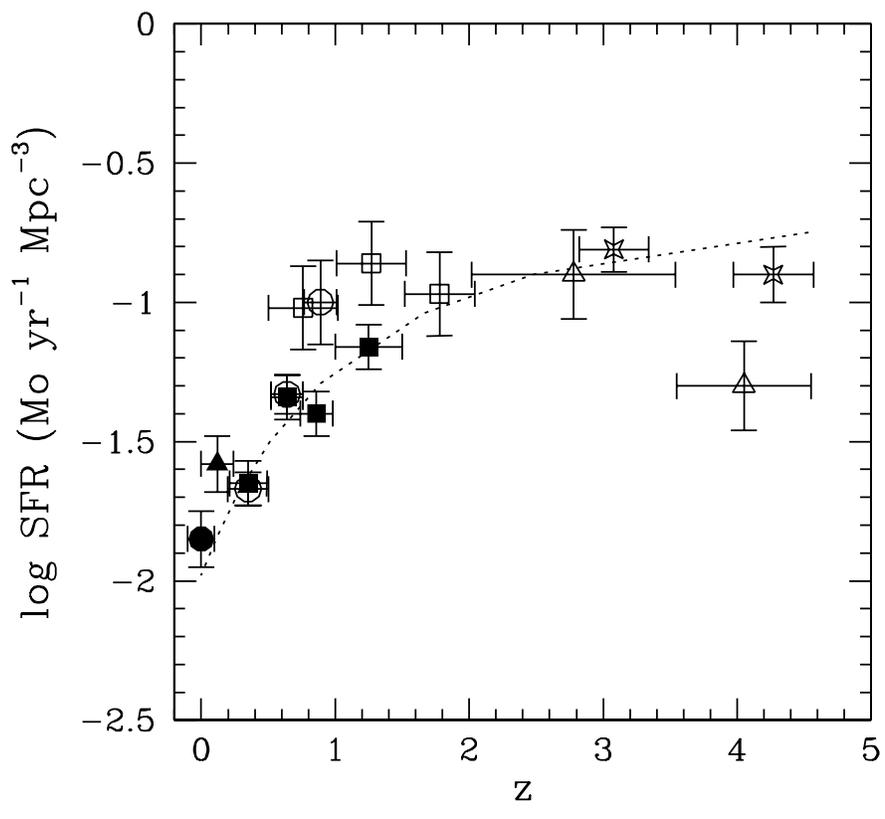
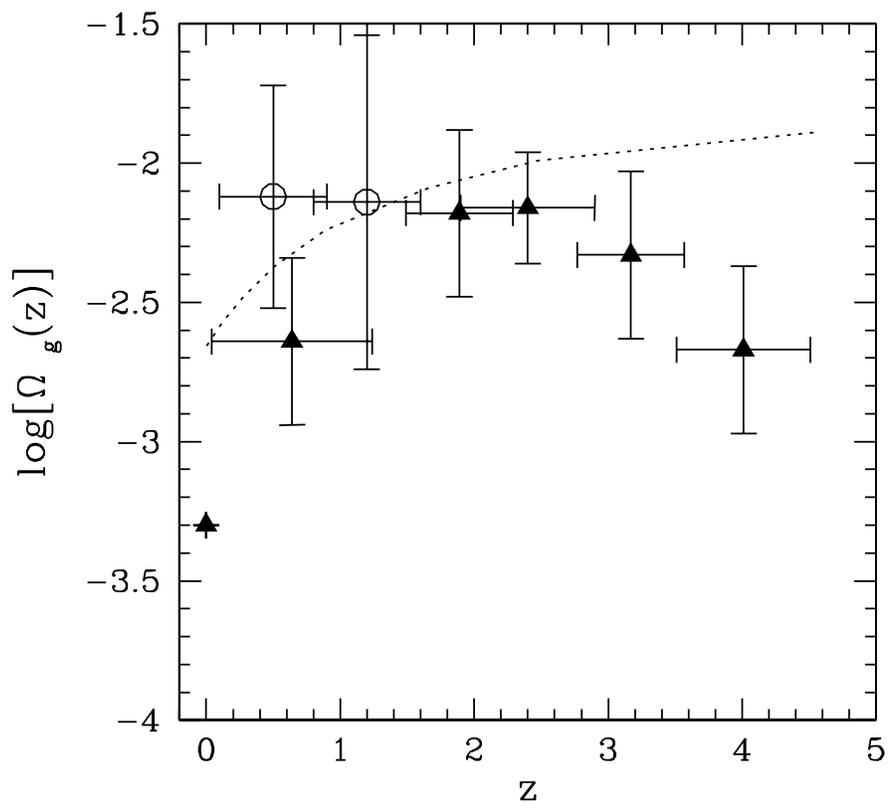
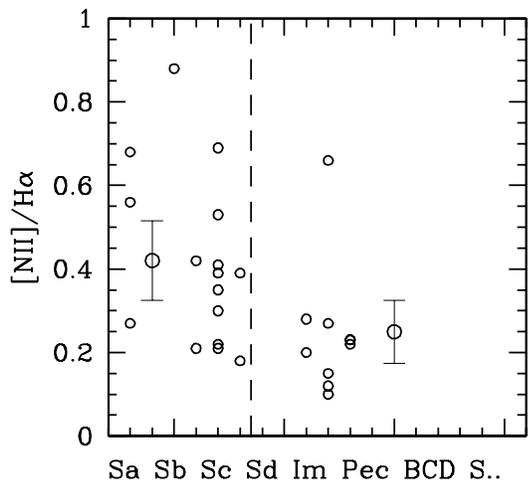
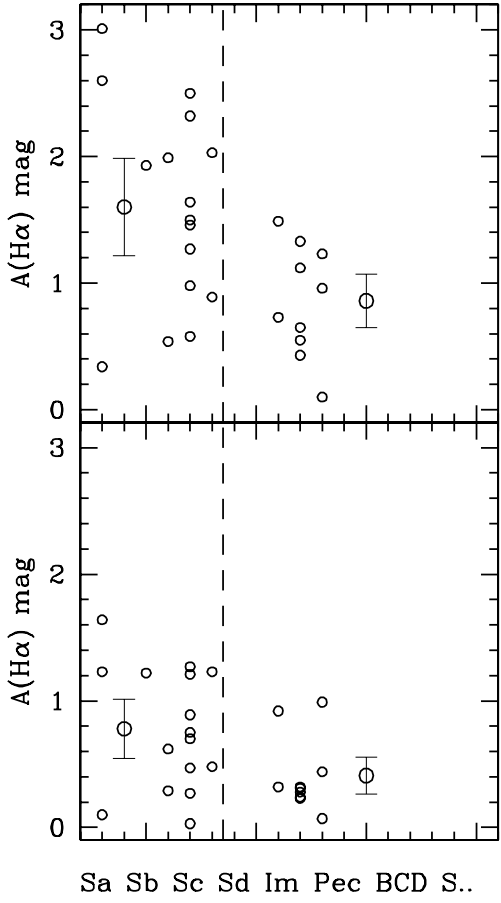
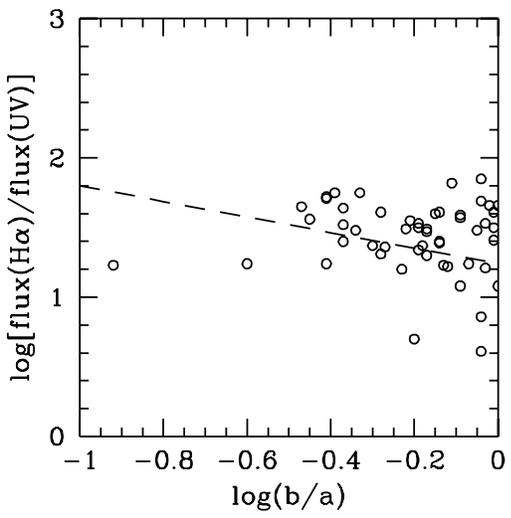
\end{document}